\documentclass[]{iscram}
\usepackage{tikz}
\usepackage{tabularx,booktabs,ragged2e}
\newcolumntype{L}{>{\RaggedRight\arraybackslash}X}

\usepackage{fancyvrb}
\usepackage{listings}
\usepackage{enumitem}
\usepackage{tcolorbox}
\usepackage{multicol}
\usepackage{siunitx}
\usepackage{xpatch}
\usepackage{multirow}
\usepackage{quoting}
\quotingsetup{vskip=1pt}
\usepackage[subrefformat=parens]{subcaption}
\usepackage{adjustbox}
\lstdefinestyle{common}{
  xleftmargin=.5em,
  xrightmargin=.5em,
  frame=single,framesep=.5em,framerule=0pt,
  fancyvrb=true,
  basicstyle=\ttfamily,
  keywordstyle=\color{cyan!50!blue!75!black}\bfseries,
  commentstyle=\color{red!50!black}\itshape,
  stringstyle=\ttfamily\color{green!50!black},
  numbers=none,
  showspaces=false,
  showstringspaces=false,
  fontadjust=true,
  keepspaces=true,
  flexiblecolumns=true,
  emphstyle=\color{red},
}
\lstdefinestyle{TeX}{
  style=common,
  backgroundcolor=\color{blue!5},
  aboveskip=5pt,
  belowskip=5pt,
  language=[LaTeX]TeX,
  moretexcs={
    abstract, addbibresource, iscramset, keywords, mainmatter,
    maketitle, printbibliography, subsection, subsubsection, url,
    urldef, href, includegraphics, ldots, parencite, citeauthor,
    citeyear, citetitle, midrule, toprule, bottomrule
  },
  fancyvrb=true,
}
\lstdefinestyle{console}{
  style=common,
  backgroundcolor=\color{gray!10},
  aboveskip=5pt,
  belowskip=5pt,
}

\newlist{options}{description}{1}
\setlist[options]{%
  beginpenalty=10000,%
  itemsep=.5\parskip plus .3\parskip minus .2\parskip,
  parsep=.5\parskip plus .3\parskip minus .2\parskip,
  topsep=.5\parskip plus .3\parskip minus .2\parskip,
  partopsep=.5\parskip plus .3\parskip minus .2\parskip,
  style=nextline,labelindent=1em,%
  font=\normalfont\ttfamily}

\colorlet{macro color}{cyan!50!blue!75!black}
\colorlet{option color}{red!50!black}
\colorlet{generic color}{green!40!black}

\newtcolorbox{pseudoTeX}{colback=blue!5,colframe=blue!5,before=\nobreak}
\let\LaTeXorig\LaTeX
\renewcommand\LaTeX{\bgroup\fontfamily{lmr}\selectfont\upshape\LaTeXorig\egroup}

\usepackage[strict]{changepage}
\usepackage{threeparttable}

\usepackage{framed}

\definecolor{formalshade}{rgb}{0.95,0.95,1}

\urldef{\sitewebgind}\url{http://gind.mines-albi.fr}
\urldef{\sitewebminesalbi}\url{http://www.mines-albi.fr}
\urldef{\gmailpaulgaborit}\url{email adress}
\urldef{\jdmail}\url{...@example.com}
\urldef{\sitex}\url{www.example.com}

\iscramset{
  CoRe Paper 2024={Social Media for Crisis Management},
  title={Monitoring Critical Infrastructure Facilities During Disasters Using Large Language Models},
  short title={Monitoring CIFs During Disasters Using LLMs},
  author={
    short name={Abdul Wahab},
    full name=Abdul Wahab Ziaullah,
    affiliation={
      Qatar Computing Research Institute, \\Hamad Bin Khalifa University, Doha, Qatar\\
       \href{mailto:awahab@hbku.edu.qa}{awahab@hbku.edu.qa}
    },
  },
  author={
    full name= Ferda Ofli,
    affiliation={
      Qatar Computing Research Institute, \\Hamad Bin Khalifa University, Doha, Qatar\\
       \href{mailto:fofli@hbku.edu.qa}{fofli@hbku.edu.qa}
    },
  },
  author={
    full name= Muhammad Imran,
    affiliation={
      Qatar Computing Research Institute, \\Hamad Bin Khalifa University, Doha, Qatar\\
       \href{mailto:mimran@hbku.edu.qa}{mimran@hbku.edu.qa}
    },
  },
}

\usepackage{tikzpagenodes}
\usetikzlibrary{fit}

\begin{document}

\maketitle


\abstract{
Critical Infrastructure Facilities (CIFs), such as healthcare and transportation facilities, are vital for the functioning of a community, especially during large-scale emergencies. In this paper, we explore a potential application of Large Language Models (LLMs) to monitor the status of CIFs affected by natural disasters through information disseminated in social media networks. 
To this end, we analyze social media data from two disaster events in two different countries to identify reported impacts to CIFs as well as their impact severity and operational status. 
We employ state-of-the-art open-source LLMs to perform computational tasks including retrieval, classification, and inference, all in a zero-shot setting. 
Through extensive experimentation, we report the results of these tasks using standard evaluation metrics and reveal insights into the strengths and weaknesses of LLMs. We note that although LLMs perform well in classification tasks, they encounter challenges with inference tasks, especially when the context/prompt is complex and lengthy. Additionally, we outline various potential directions for future exploration that can be beneficial during the initial adoption phase of LLMs for disaster response tasks.

}


\section{Introduction}
\label{sec:intro}

Critical infrastructures, including essential facilities and services vital for societal functioning, span sectors like energy, transport, healthcare, and telecommunications~\parencite{pescaroli2017critical, labaka2016holistic}. Ensuring the continuous functioning and accessibility of critical infrastructures is essential in assisting vulnerable populations during disasters and  
disruptions to these critical infrastructures can result in increased human and economic losses \parencite{auerswald2005challenge}.
However, during major disasters, authorities are overwhelmed and thus face challenges in maintaining an updated status of these facilities, and as a consequence, the general public is kept uninformed about the latest status of these facilities. 

This paper bridges the gap by exploring an unconventional data source---social media---to monitor Critical Infrastructure Facilities (CIFs) in a given area of interest (AOI). Social media platforms, especially microblogging sites such as X (formerly Twitter), are important data sources for real-time disaster updates. People post information about early warnings, cautions, damages, and their needs almost instantaneously~\parencite{olteanu2015expect, qu2011microblogging}. Past studies show the use of social media information for several disaster response tasks including damage and needs assessment, among others~\parencite{imran2015processing,castillo2016big}.

While social media holds valuable information, it frequently becomes overloaded with noise—comprising casual chatter filled with slangs, abbreviations, and grammatically unconventional sentences. Past studies suggest employing supervised classification techniques to discern social media messages relevant to disaster response~\parencite{imran2015processing}. However, these techniques require human-labeled data for each unique classification task or even when adapting a model trained on past disaster data to a new emergency. For example, \textcite{alam2021humaid} manually labeled around 77K tweets to train deep neural network models such as BERT, RoBERTa, for detecting situational awareness messages. Similarly, \textcite{zahra2020automatic} shows an extensive feature engineering process to train classifiers for eyewitness message detection. These methods are tailored for specific classification tasks, and should a new task/class be needed, retraining of the models would be required. To tackle this challenge, our work explores the use of Large Language Models (LLMs) to perform traditional computational tasks, an effort toward eliminating the requirement for explicit model training.

To monitor CIFs in a specified AOI, our proposed approach starts by acquiring all critical facilities in the area from Open Street Maps (OSM) through automatic APIs. Social media data collected about an ongoing disaster is processed through an LLM to generate embeddings and stored in a vector database.\footnote{The list of CIFs, data, and embeddings are available upon request.} Embedding of a message encapsulates its semantics and thus yields better search results. Therefore, for each CIF obtained from OSM, we query the vector database, where raw embeddings are stored, to retrieve messages pertinent to the specified CIF. The retrieved messages are then analyzed by LLMs to identify \textit{(i)} the impacts reported in each message, \textit{(ii)} the severity of the reported impacts, and \textit{(iii)} the operational status of the CIF. We perform all the tasks in a zero-shot setting and evaluate each step of the proposed methodology using standard evaluation metrics.

Through extensive experimentation, we show both the strengths and limitations of LLMs in handling various computational tasks associated with the processing of social media messages. Notably, we showcase diverse configurations of our CIF retrieval query and the corresponding advantages they offer. For instance, our CIF query consisting of a fixed general term (e.g., ``disaster impacts") yields better outcomes compared with a detailed listing of impacts. Furthermore, we note better classification performance of LLMs for individual messages when compared to the classification and inference from a set of messages. Overall, we demonstrate that LLMs, even in the zero-shot setting, hold great potential to replace traditional supervised models. However, there are certain weaknesses as well. Notably, LLMs may struggle with context understanding and misinterpret nuanced or ambiguous language commonly found in social media. Additionally, their performance can be influenced by factors like data biases and the nuances of prompt engineering. Acknowledging these limitations is crucial to refining the application of LLMs in processing social media messages effectively. 

The rest of the paper is structured as follows. The following section reviews Related Work. In the Methodology section, we provide details of our proposed approach. Results are presented in the Results section and further deliberated upon in the Discussion section. The paper concludes in the final section.

\section{Related Work}
\label{sec:related_work}


Critical infrastructures are at the heart of economic and social welfare in all countries. Therefore, understanding their risk and resilience to various hazards and disasters have been crucially studied in literature \parencite{yusta2011methodologies}. For example, \textcite{fekete2015critical} explored the use of Geographic Information Systems (GIS) and Remote Sensing (RS) whereas \textcite{jovanovic2018indicator} presented an indicator-based approach to assess resilience of critical infrastructures to disasters. To this end, \textcite{baloye2016modelling} proposed a critical infrastructure-driven spatial database for proactive disaster management. Taking a step further, \textcite{qiang2019flood} provided a comprehensive assessment of flood exposure of all major critical infrastructures in the United States. 


In the last decade, with the advancement of social networks, researchers started to explore the use of real-time social media data during disasters to assess risk and resilience of critical infrastructures. For instance, \textcite{mittelstadt2015integrated} integrated social media data with mobile in-situ data to monitor the disaster impact on critical infrastructure. 
Later \textcite{fan2019graph} developed a graph-based method for detecting credible situation information related to infrastructure disruptions during natural disasters. 
In another study, \textcite{roy2020multilabel} presented a multilabel classification approach to identify hurricane-induced infrastructure disruptions through a sentiment analysis of social media data. They utilized several supervised machine learning models to classify disruptions in social media messages. 
Alternatively, \textcite{heglund2021social} demonstrated the utility of social media data for critical infrastructure resilience based on statistical analysis and forecasting methods. While plethora of these methods utilize conventional pipelines of message filtration, trend detection, and other case-specific algorithms, our methodology leverages more generalized inferential capabilities of the large language models (LLMs). 

With the emergence of pre-trained LLMs such as GPT3.5 \parencite{brown2020language}, GPT4 \parencite{openai2023gpt4}, Llama2 \parencite{touvron2023llama}, Mistral 7B \parencite{jiang2023mistral}, more versatile capabilities have become available for research. For instance, \textcite{colverd2023floodbrain} has recently studied the use of LLMs for generating flood disaster impact reports via a system initiated by a search query phrase, which is then utilized to gather textual data about the disaster from the web. 
Another recent use of LLM is geotagging of tweets where few-shot learning is utilized for identification of various location descriptors in tweets such as door number addresses, road segments, and road intersections \parencite{hu2023geo}. 

Inspired by this trend, we capitalize on an LLM for real-time monitoring of critical infrastructure facilities (CIFs) during disasters by classifying their situations in three crucial aspects including the type of impact, its severity, and the operational status. Additionally, to compensate for lack of a publicly available dataset with situational information about CIFs, we again revert to an LLM for its generative capabilities to create synthetic data to test our CIF monitoring pipeline.

\section{Methodology}
\label{sec:method}

\begin{figure}[t]
  \centering
  \includegraphics[width=\linewidth]{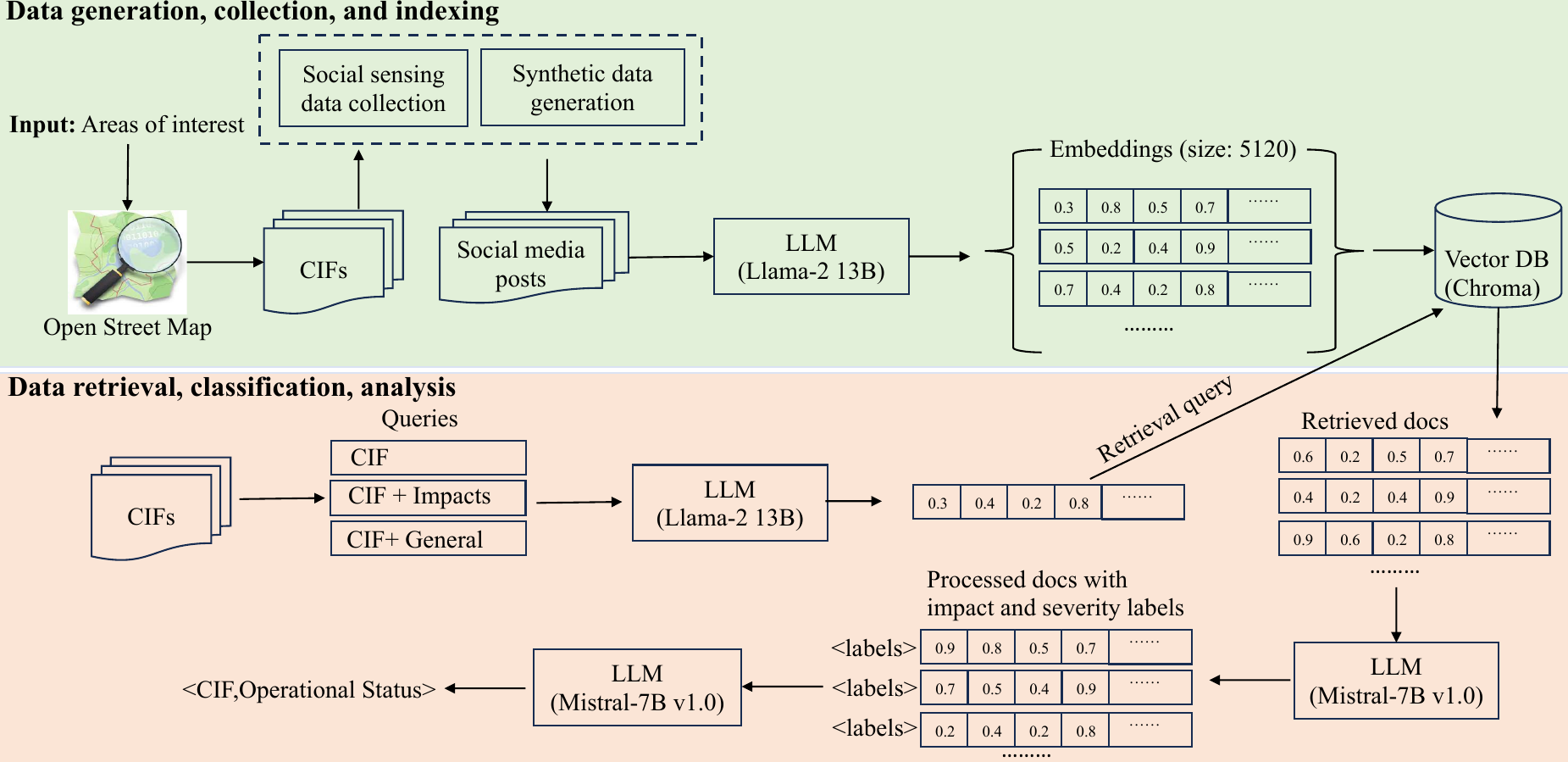} 
  \caption{High-level methodology detailing two pipelines: \emph{(i)} data generation, collection, and indexing, \emph{(ii)} data retrieval, classification, and analysis}
  \label{fig:method}
\end{figure}


\subsection{Overview}
The primary objective of this work is to perform real-time monitoring of various Critical Infrastructure Facilities (CIFs) amidst large-scale disaster scenarios. Specifically, we aim to obtain three types of updates for each CIF in the given area of interest, including \emph{(i)} identifying the disaster impact (e.g., damaged, burned) to the CIF, \emph{(ii)} assessing the severity of the impact (e.g., moderate, severe) and \emph{(iii)} determining the operational status of the facility (e.g., closed, partially closed, or open). We employ social sensing data from social media platforms, harnessing its capability to potentially deliver immediate updates, potentially from eyewitnesses or people possessing any relevant information.

Figure~\ref{fig:method} depicts the high-level methodology comprising two pipelines: \emph{(i)} data collection and indexing, and \emph{(ii)} data retrieval and processing. The data collection and indexing pipeline performs tasks related to data acquisition, synthetic data generation, embedding generation, and indexing. The pipeline for data retrieval and processing focuses on retrieving pertinent data from the embedding database. Subsequently, the retrieved data are analyzed through LLMs to identify the impact, severity, and operational status of CIFs. Next, we provide detailed descriptions of the important processes of the methodology.

\subsection{Obtaining CIFs}
We utilize Nominatim APIs\footnote{https://nominatim.openstreetmap.org/} from Open Street Map (OSM)~\parencite{OpenStreetMap} to retrieve information about CIFs within a specified geographical area of interest (AOI). In this work, we focus on two regions prone to natural disasters: Christchurch, New Zealand, susceptible to earthquakes, and Broward County, Florida, USA, susceptible to hurricanes. We note that for both AOIs we also have Twitter data previously collected from these locations during past disasters. To extract CIF data for these AOIs, we construct queries that include the AOI name and the type of infrastructure, including hospitals, fire stations, rail stations, educational facilities, bridges, roads, and tunnels. Example queries include ``Christchurch hospitals,'' ``Christchurch fire stations,'' etc. Employing this approach, we gathered 58 CIFs from Christchurch and 82 CIFs from Broward County, along with associated metadata such as names, addresses, and geographic coordinates. 

\subsection{Data Curation: Real and Synthetic}
Social media data about CIFs is central to our study. However, to the best of our knowledge, there is no publicly available social media data containing tagged CIFs and their associated impacts, severity, and operational statuses. To address this gap, we leverage generative AI to create synthetic data (i.e., tweets) that include real-world CIFs from our selected AOIs. In particular, we utilize the open-source Llama-2 13B model~\parencite{touvron2023llama} to generate tweets reporting diverse impact scenarios on CIFs retrieved from OSM. Moreover, for each generated tweet, we ask the model to provide the reported impact along with its corresponding severity labels. The specific prompt employed for tweet generation for Broward County is as follows:

\begin{quoting}
{\small{\it{Generate 15 diverse tweets describing the impact of a Category-5 hurricane on [CIF NAME \& ADDRESS]. The disaster triggered sub-events such as tornado, storm surge, burst of rain, strong winds, resulting in varied impacts like flooded, collapsed, submerged, damaged, destroyed, cracked, etc.

Ensure linguistic diversity in each tweet, providing unique insights into the impact and its severity. In some tweets, include the infrastructure's address. Aim for tweet lengths between 100 to 250 characters, and avoid using emojis. Tag each tweet with the type of impact (e.g., damaged, destroyed) and its severity (e.g., low, mild, severe). These tags will be used for training classifiers.

Always include 2 tags at the `end' of the generated tweet with the following template:
(Tags: ***** , *****)
}}}
\end{quoting}

For instance, the response returned by the LLM for CIF \textit{``Aventura Hospital and Medical Center \& 20900, Biscayne Boulevard, Aventura, Miami-Dade County, Florida, 33180, United States''} looks as follows (showing first three tweets for brevity):

\begin{quoting}
{\small{\it{
Aventura Hospital's emergency room is flooded after a burst of rain caused by the Category 5 hurricane. Patients and staff are being evacuated to safer areas. (Tags: Flooded, Mild)

Strong winds from the hurricane have caused significant damage to Aventura Hospital's roof and windows, leaving many areas closed off to patients and staff. (Tags: Damaged, Severe)

The storm surge from the hurricane has submerged the lower levels of Aventura Hospital, forcing patients and staff to be evacuated to higher floors. (Tags: Submerged, Severe)
}}}
\end{quoting}

A similar prompt is used for Christchurch where the first paragraph is adapted for an earthquake scenario as follows:

\begin{quoting}
{\small{\it{Generate 15 diverse tweets describing the impact of a severe earthquake on Christchruch's [CIF NAME \& ADDRESS]. The disaster triggered sub-events such as ground shake, landslide, liquefaction, ground rupture, aftershock, resulting in varied impacts like collapsed, cracked, damaged, destroyed, etc.
}}}
\end{quoting}


Subsequently, we utilize the same LLM employed for tweet generation to assign an operational status label to each generated tweet. The model is restricted to select one of four operational statuses: ``open,'' ``closed,'' ``partially open,'' ``partially closed,'' and ``unknown" if no operational status is explicitly mentioned in the tweet. The prompt to obtain the operational status is as follows:

\begin{quoting}
{\small{\it{Your task is to analyze the provided tweet and determine the operational status of the mentioned infrastructure. The operational status could include descriptors such as open, closed, partially open, partially closed, or unknown. 

Tweet: [TWEET]

Operational status: 
}}}
\end{quoting}

A sample response from the above prompt for the tweet \textit{``The computer lab at Christchurch Girls' High School has been closed due to a collapsed ceiling. Students are using temporary facilities until further notice. \#ChristchurchEarthquake''} is as follows:
\begin{quoting}
{\small{\it{
Operational status: closed
}}}
\end{quoting}

\begin{figure}[t]%
    \centering
    \begin{subfigure}[b]{.48\linewidth}
    \centering
    \includegraphics[width=\linewidth]{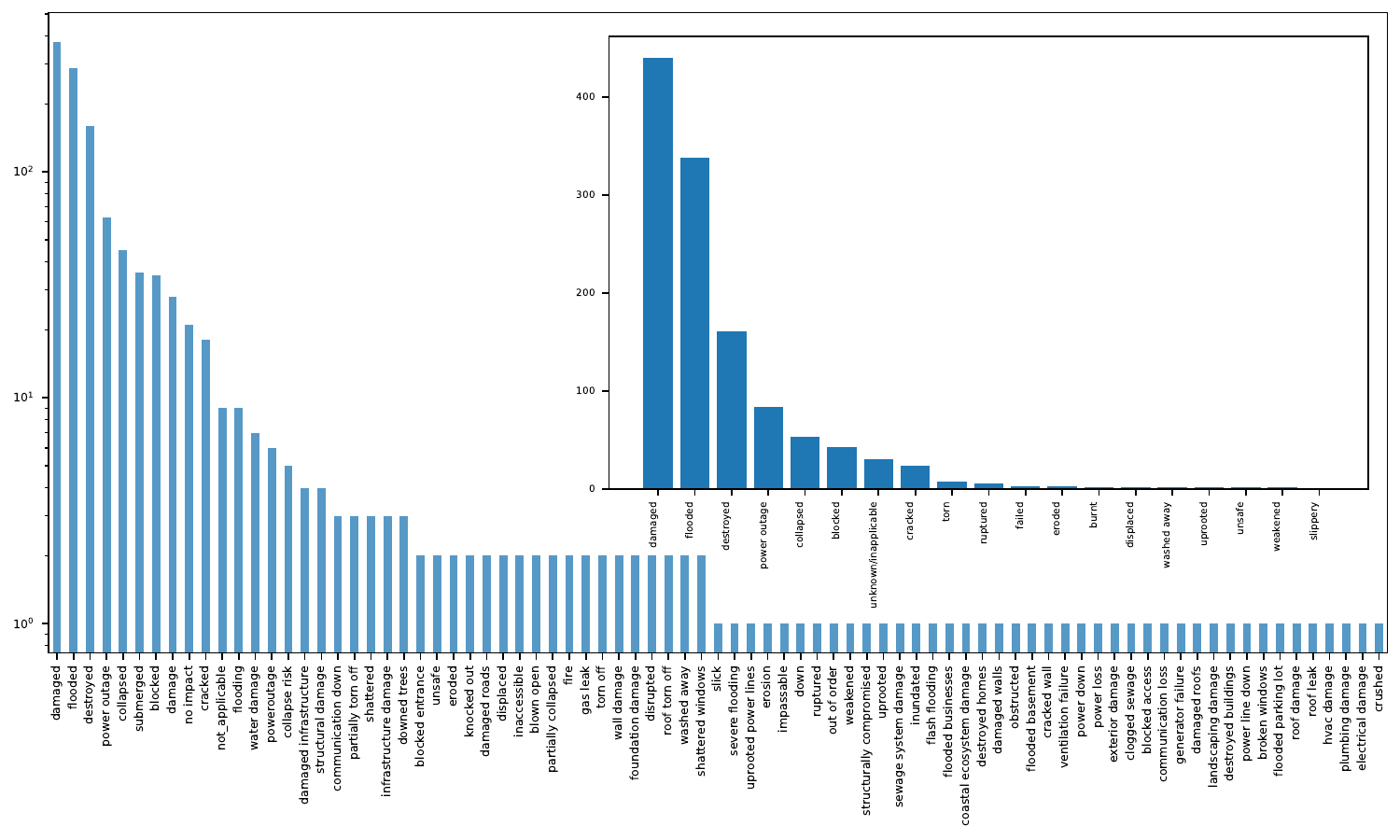}
    \caption{Broward County}
    \end{subfigure}
    \hfill
    \begin{subfigure}[b]{.48\linewidth}
    \centering
    \includegraphics[width=\linewidth]{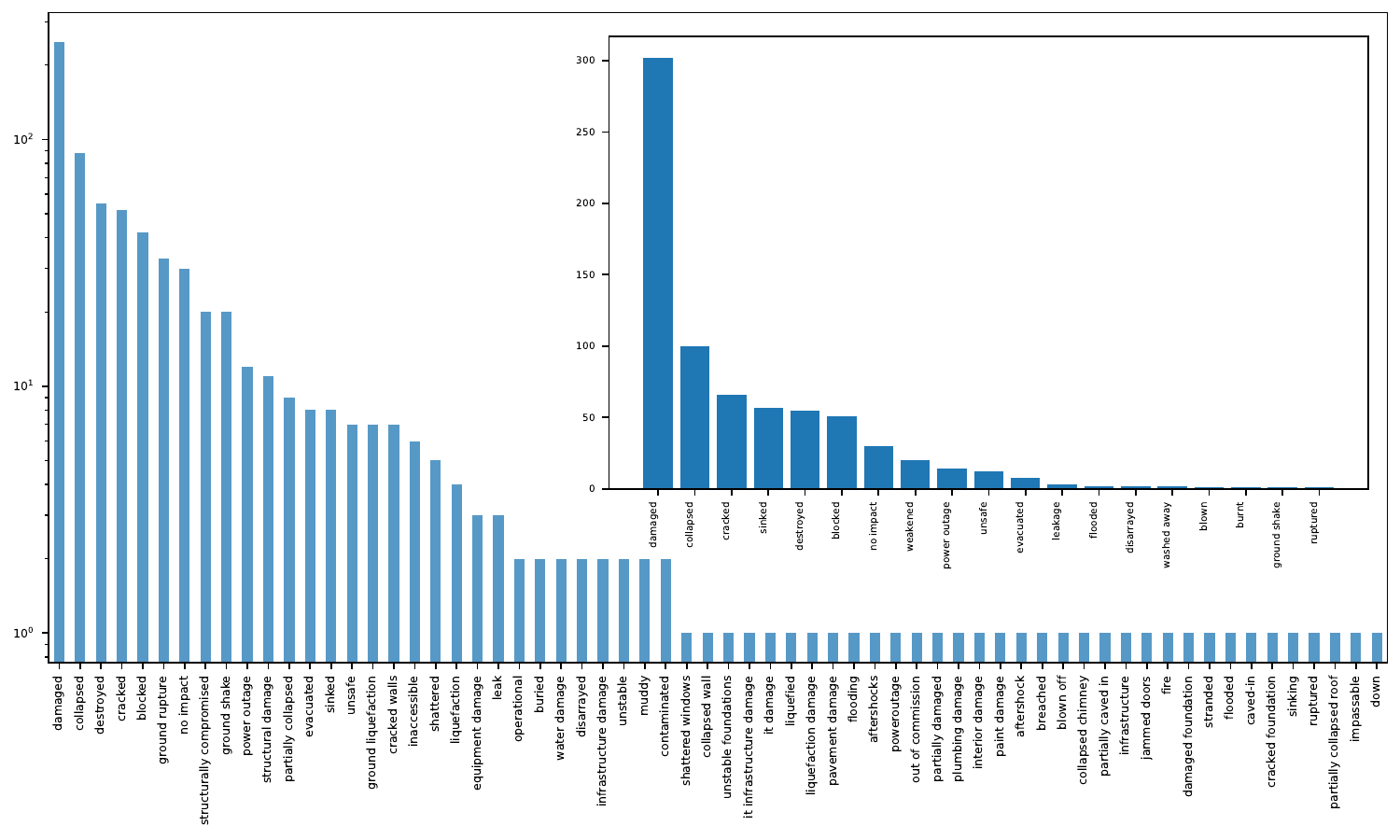}
    \caption{Christchurch}
    \end{subfigure}
    \caption{Distribution of impact labels in the synthetic data for (a) Broward County and (b) Christchurch. The outer charts with light blue bars correspond to the LLM-generated raw tags whereas the overlaid smaller charts with dark blue bars show the manually pruned ground-truth tags.}
    \label{fig:impact_frequency_distribution}%
\end{figure}

In total, we generated 728 tweets for Christchurch for 58 CIFs and 1,205 tweets for Broward County against 82 CIFs. Although instructed, the model does not always generate 15 tweets for each CIF. Tweets with incorrect tags/labels were manually corrected. As we did not control the prompt to enforce a closed-set of impact types, the resulting list of impact tags is open-ended (N=83), which also allowed for more diversity in the generated tweets. Figure~\ref{fig:impact_frequency_distribution} shows the distribution of the resulting tags for each AOI. 
We noticed multiple similar impact labels (e.g., flooded, flooding, inundated, etc.) in the model responses, which we merged into a unified list of tags, as illustrated in Table~\ref{table:impact_mappings}. The sub-figures in Figure~\ref{fig:impact_frequency_distribution} depict the distribution of the refined tag list.
Note that, unlike impact tags, we utilized a closed-set of tags for classification of severity (i.e., severe, moderate, mild, and unknown) and operational status (i.e., open, closed, partially open, partially closed, and unknown).


\begin{table}[t]
\caption{Raw impact labels and their mapping to final impact taxonomy for LLM-based classification}
\label{table:impact_mappings}
\setlength\tabcolsep{3pt} 
\centering
\begin{adjustbox}{max width=\linewidth}
\begin{tabularx}{\textwidth}{lX}
\toprule
Consolidated labels   & Raw model labels\\
\midrule
damaged              & aftershock damage, damage, damaged, broken windows, damaged infrastructure, damaged roads, damaged roofs, damaged walls, electrical damage, coastal ecosystem damage, foundation damage, landscaping damage, exterior damage, hvac damage, infrastructure damage, roof damage, plumbing damage, sewage system damage, structural damage, wall damage, water damage, equipment damage, damaged foundation, interior damage, IT damage, liquefaction damage, partially damaged, pavement damage, plumbing damage, IT infrastructure damage, equipment damage, paint damage \\
flooded              & flooded basement, flooded businesses, flooded parking lot, flooding, severe flooding, flash flooding, submerged, inundated\\
destroyed            & destroyed, destroyed buildings, destroyed homes, rubble \\
weakened             & weakened, structurally compromised \\
cracked              & cracked, shattered, shattered windows, debris, cracked wall, cracked walls \\
blocked              & blocked access, blocked entrance, impassable, obstructed, clogged sewage, inaccessible, partially blocked, landslide, jammed doors \\
torn                 & partially torn off, roof torn off \\
power outage         & power line down, power loss, power outage, power down, communication down, communication loss, down, knocked out, disrupted, downed trees, offline \\
ruptured             & blown open, gas leak, roof leak \\
collapsed            & partially collapsed, collapsed, crushed, collapse risk, collapsed wall, collapsed chimney, partially collapsed roof \\
failed               & generator failure, ventilation failure, out of order \\
uprooted             & uprooted, uprooted power lines \\
eroded               & eroded, erosion \\
washed away          & washed away, muddy \\
slippery             & slippery, slick \\
displaced            & displaced \\
blown                & blown, blown off \\
burnt                & burnt, burning, fire \\
unsafe               & unsafe, unstable, structurally compromised, uninhabitable, contaminated \\
leakage              & leak, gas leak, gasleak \\
sinked               & ground liquefaction, liquefaction, buried, sinked, sinking, caved-in, ground rupture, liquefied \\
unknown/inapplicable & not applicable, not humanitarian, no impact, unknown \\
\bottomrule
\end{tabularx}
\end{adjustbox}
\end{table}

Later, for each CIF, we dispersed its generated tweets across four six-hour time intervals (spanning a 24-hour time frame) by considering their impact order (e.g., ``destroyed" followed by ``damaged" or ``cracked" and so on). The impact ordering ensures a coherent progression of the impact to a CIF. 
This step is crucial for simulating a real-world setting in our subsequent experiments. Moreover, given that several past studies highlight the prevalence of noise and the limited relevant information in real-world tweets \parencite{kumar2017authenticity,grace2021toponym}, we aim to ensure the realism of our dataset. To achieve this, we introduce noise into our generated data with a signal-to-noise ratio of 2\% (signal) to 98\% (noise). 
For this purpose, we incorporated tweets collected during past disasters for both AOIs. Specifically, for Christchurch, we used tweets collected after a magnitude-7.1 earthquake struck New Zealand on September 2nd, 2016. For Broward County, we utilized tweets collected after Hurricane Ian hit Florida in September 2022. In total, we prepared 36,286 and 60,062 tweets (including both synthetic and real) for Christchurch and Broward County, respectively. To ensure that CIFs also appear in non-impact tweets, we injected CIF names into 8\% of the noisy tweets in each time interval.

Finally, to assign an overall status for a CIF within each time interval, we deduce its operational status. To achieve this, we chronologically order tweets for a CIF and use the operational status of the last tweet as an overall status. If the operational status of the last tweet is ``unknown", we use the last valid status.

\subsection{Generating and Indexing Tweet Embeddings}
Next, we employ Llama-2 13B~\parencite{touvron2023llama} to compute embeddings for all the tweets, including synthetic and real. Each tweet is converted into a 5120-dimensional embedding vector which is the default embedding size of Llama-2 13B. This embedding is then indexed into ChromaDB\footnote{\url{https://www.trychroma.com/}}, a well-known open-source vector database. These embeddings serve as the basis for semantic search, enabling the retrieval of tweets specifically mentioning certain CIFs.

\subsection{Retrieval of CIF Tweets}

The first step in determining the impact and operational status of a CIF involves retrieving tweets referencing a specific CIF. To achieve this, we explore three types of retrieval queries: \textit{(i)} a query comprising only the CIF name, \textit{(ii)} a query involving both the CIF name and a set of impact terms such as damaged, cracked, burned, destroyed, submerged, and \textit{(iii)} a query combining the CIF name with the term ``disaster impacts.'' For each query, we first obtain its embeddings from the Llama-2 13B model and subsequently retrieve the top-$K$ (ranging from 5 to 50 with a step size of 5) tweets by querying ChromaDB using cosine similarity.

Top-$K$ results for each query are then used to compute Average Precision (AP) using Equation (\ref{AP_eq}).

 

\begin{align}
\mathrm{AP}@K & =\frac{1}{\mathrm{GTP} @ K} \sum_{\substack{k=5n\\1\leq n \leq N}}^{} \mathrm{P} @ k \times \mathrm{rel} @ k \text{, for } K=5N \text{ and } N=\{1,\ldots,10\}
\label{AP_eq}
\end{align}
where
\begin{align*}
\mathrm{P} @ k & =\frac{\text{Number of relevant documents retrieved } @ k }{ k }\\
\mathrm{rel} @ k & = \begin{cases} 1, & \text{if $k^{th}$ document is relevant} \\ 0, & \text{otherwise} \end{cases} \\
\mathrm{GTP} @ K &= \min(K,\text{Total number of relevant documents})
\end{align*}

In Equation (\ref{AP_eq}), we normalize the total sum by the number of `retrievable' ground-truth documents or \textit{signals} (i.e., Ground Truth Positive or GTP). Finally, to determine the overall retrieval performance of a given query, we compute mean Average Precision (i.e., mAP) in a time interval $t \in \{\text{0h-6h}, \text{6h-12h}, \text{12h-18h}, \text{18h-24h}, \text{0-24h}\}$ using Equation (\ref{MAP_eq}).


\begin{align}
\mathrm{mAP}@K^{t}=\frac{1}{M} \sum_{i=1}^{M} \mathrm{AP}@K_{i}^{t} \label{MAP_eq} \text{, for } K=5N \text{ and } N=\{1,\ldots,10\}
\end{align}


\subsection{Tweet Classification for Impact, Severity, and Operational Status}
We employ the Mistral 7B v1.0 model~\parencite{jiang2023mistral} for classifying the retrieved tweets that are identified as relevant by the CIF queries. These tweets are likely to encompass information about impacts experienced by critical facilities. To perform the impact and severity classification of each tweet, we use the following prompt template:

\begin{quoting}
{\small{\it{Your task is to analyze the provided tweet and determine the impacts of the disaster on the mentioned infrastructure. Please include only impact descriptors from the list such as blocked, blown, buried, burnt, collapsed, cracked, damaged, destroyed, displaced, disrupted, eroded, failed, flooded, ground liquefaction, ground shake, leakage, muddy, power outage, ruptured, slippery, torn, unsafe, uprooted, washed away, weakened or not\_applicable, and severity such as severe, mild, moderate, unknown. 

Tweet: [TWEET]

Infrastructure impact: \\
Infrastructure severity: 
}}}
\end{quoting}

The response returned by the model to an example tweet such as \textit{``The Christchurch Public Hospital's radiology equipment is malfunctioning due to the earthquake, making it difficult to diagnose patients. \#ChristchurchEarthquake''} looks as follows:

\begin{quoting}
{\small{\it{Infrastructure impact: damaged\\
Infrastructure severity: moderate}}}
\end{quoting}

Finally, tweets that are tagged with an impact type are further analyzed to infer the operational status of CIFs. For this purpose, we use the Mistral 7B v1.0 model with the following prompt:

\begin{quoting}
{\small{\it{Your task is to analyze the provided tweet and determine the operational status of the mentioned infrastructure. The operational status could include descriptors such as open, closed, partially open, partially closed, or unknown. 
    
Tweet: [TWEET]\\
Operational status:
}}}
\end{quoting}

The response returned by Mistral 7B v1.0 model to an example tweet such as \textit{``Ground shake from Christchurch earthquake caused significant damage to Wigram Fire Station's foundation, rendering it unstable. \#ChristchurchEarthquake''} looks as follows:

\begin{quoting}
{\small{\it{Operational status: closed}}}
\end{quoting}

To evaluate the performance of impact, severity, and operational status classification, we compare the labels predicted by the Mistral 7B model with the ground-truth labels obtained from the Llama-2 model and report results using standard evaluation metrics such as precision, recall, and F1-score.

\subsection{Obtaining Overall Operational Status of CIFs}


The last step in the tweet classification and analysis pipeline involves extracting the overall operational status of tweets retrieved through a CIF query. The set of tweets resulting from a CIF query may encompass information about the queried CIF, but these tweets may vary in terms of containing impact-related details. Therefore, to determine the overall operational status of a CIF at a specific time, we only consider tweets containing impacts (relying on the impact labels obtained in the previous step). To achieve this, we employ the Mistral 7B v1.0 model with the following prompt to retrieve the overall status. 


\begin{quoting}
{\small{\it{Your task is to analyze the tweets given below and deduce the operational status of a facility, named [CIF]. Since these tweets are retrieved based on the facility name, it's possible that some tweets may not pertain to the given facility. Focus solely on the tweets pertinent to [CIF] and derive the most recent operational status for the facility. Your operational status label must be one of these: open, closed, partially open, partially closed, or unknown. 

Tweet: [TWEET 1]\\
Tweet: [TWEET 2]\\
Tweet: [TWEET 3]\\
...

operational\_status: 
}}}
\end{quoting}

\section{Results}

In this section we present our results for the retrieval and classification tasks described in the Methodology section.

\subsection{Retrieval of CIF Tweets}

For each AOI, we considered three types of queries to fetch relevant tweets pertaining to each CIF:
\begin{enumerate}[label=(\roman*),itemsep=0mm]
    \item \textit{CIF} --- a query comprising only the CIF name
    \item \textit{CIF + X} --- a query involving both the CIF name and a set of impact terms such as damaged, cracked, burned, destroyed, submerged, etc. (denoted as X for brevity)
    \item \textit{CIF + disaster impacts} --- a query combining the CIF name with the term ``disaster impacts''
\end{enumerate}
Specifically, for Broward County \textit{X} comprises \textit{flooded, submerged, damaged, destroyed, weakened, cracked, blocked, torn, power outage, ruptured, collapsed, failed, uprooted, eroded, burnt, washed away, slippery, displaced, disrupted} whereas for Christchurch it consists of \textit{flooded, destroyed, leak, blocked, cracked, ground liquefaction, power outage, ruptured, buried, collapsed, ground shake, unsafe, muddy}.

\begin{figure}[t]
    \centering
    \begin{subfigure}[b]{.19\linewidth}
    \centering
    \includegraphics[width=\linewidth]{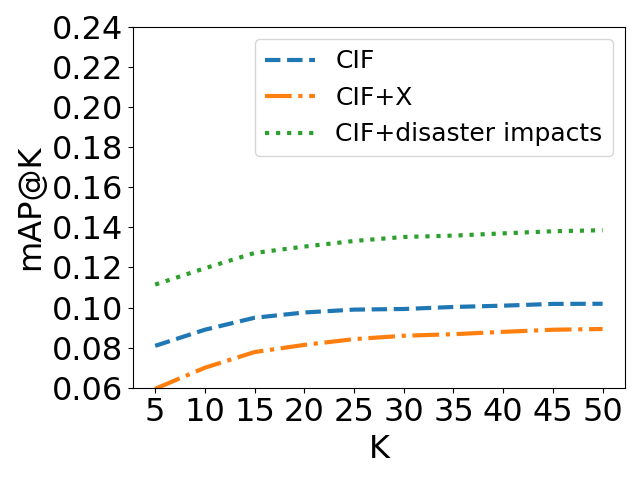}
    \caption{0h-6h}
    \end{subfigure}
    \begin{subfigure}[b]{.19\linewidth}
    \centering
    \includegraphics[width=\linewidth]{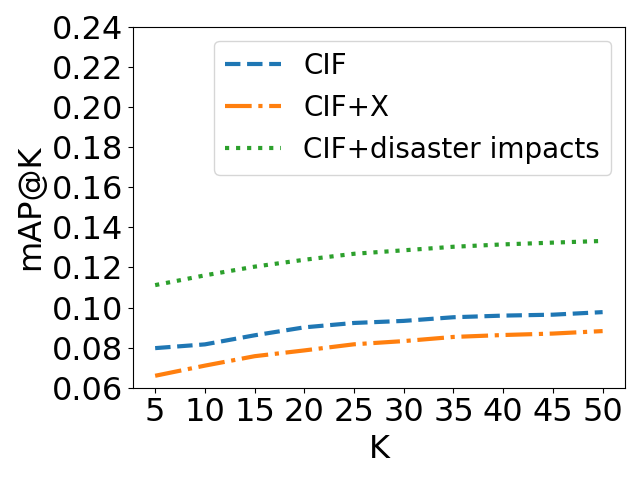}
    \caption{6h-12h}
    \end{subfigure}
    \begin{subfigure}[b]{.19\linewidth}
    \centering
    \includegraphics[width=\linewidth]{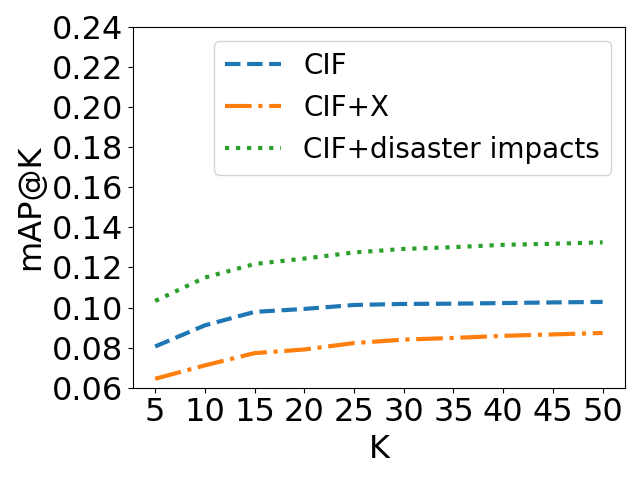}
    \caption{12h-18h}
    \end{subfigure}
    \begin{subfigure}[b]{.19\linewidth}
    \centering
    \includegraphics[width=\linewidth]{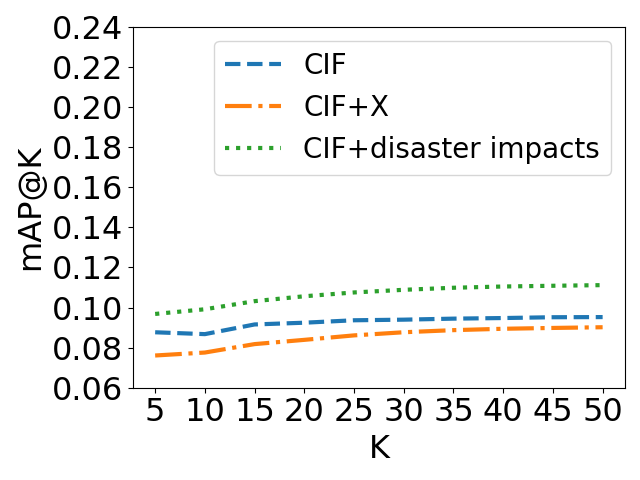}
    \caption{18h-24h}
    \end{subfigure}
    \begin{subfigure}[b]{.19\linewidth}
    \centering
    \includegraphics[width=\linewidth]{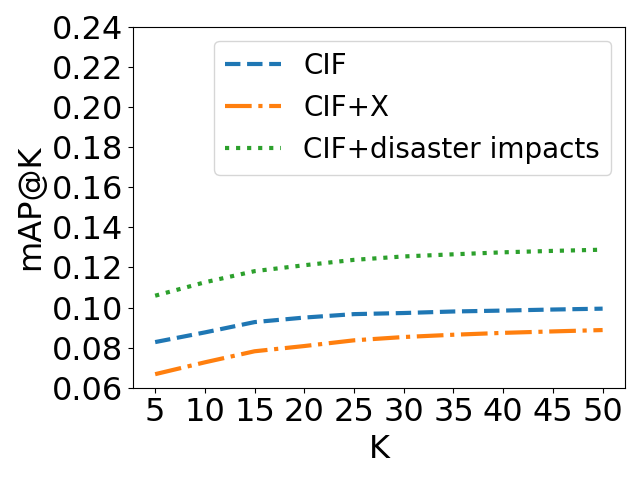}
    \caption{0h-24h}
    \end{subfigure}
    \caption{Performance comparison of retrieval queries for Broward County}
    \label{fig:broward_retrieval_mAP}%
\end{figure}

\begin{figure}[t]
    \centering
    \begin{subfigure}[b]{.19\linewidth}
    \centering
    \includegraphics[width=\linewidth]{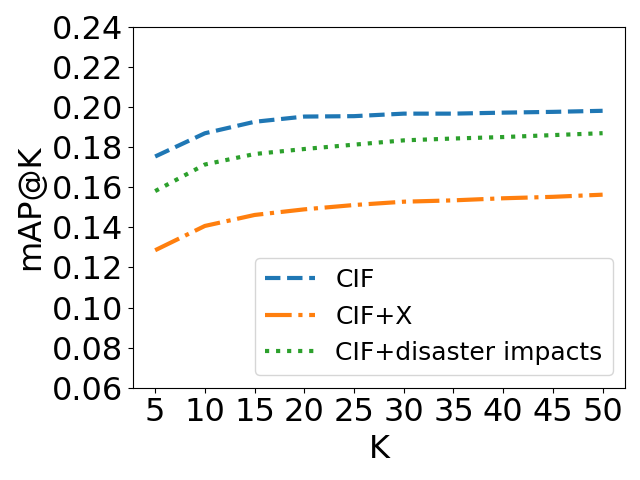}
    \caption{0h-6h}
    \end{subfigure}
    \begin{subfigure}[b]{.19\linewidth}
    \centering
    \includegraphics[width=\linewidth]{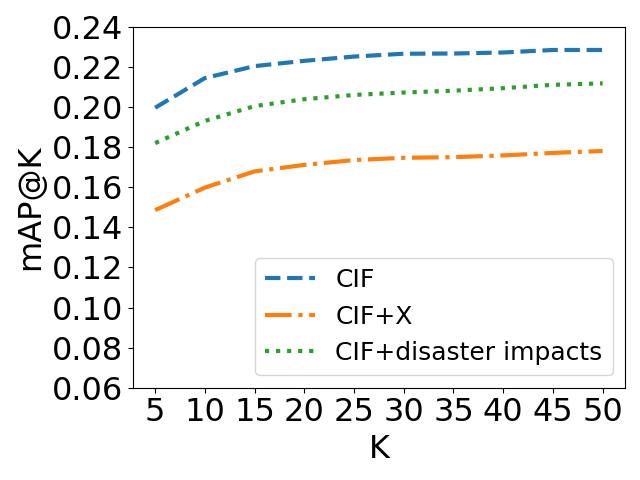}
    \caption{6h-12h}
    \end{subfigure}
    \begin{subfigure}[b]{.19\linewidth}
    \centering
    \includegraphics[width=\linewidth]{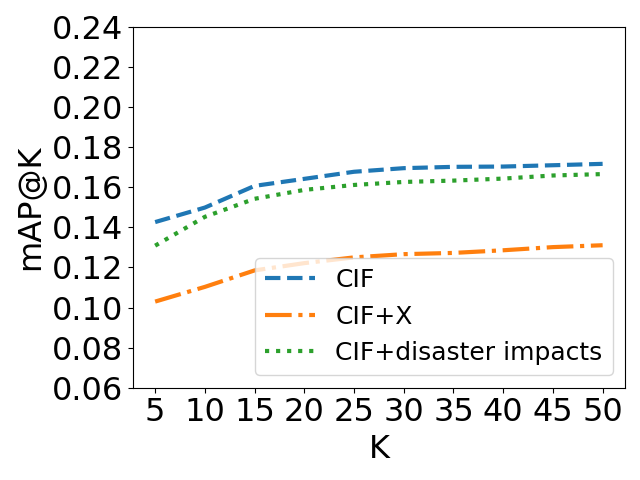}
    \caption{12h-18h}
    \end{subfigure}
    \begin{subfigure}[b]{.19\linewidth}
    \centering
    \includegraphics[width=\linewidth]{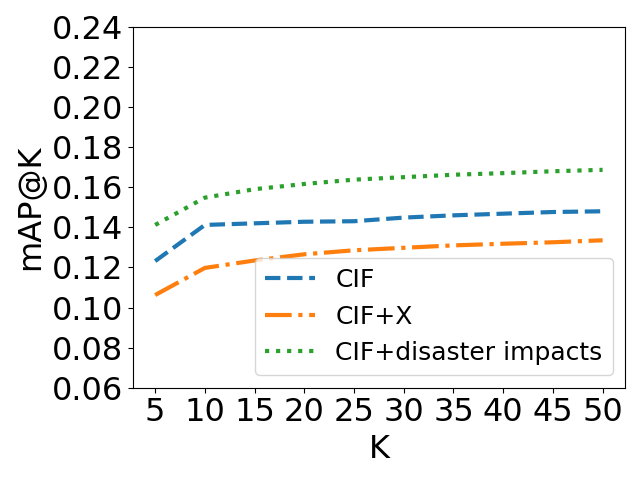}
    \caption{18h-24h}
    \end{subfigure}
    \begin{subfigure}[b]{.19\linewidth}
    \centering
    \includegraphics[width=\linewidth]{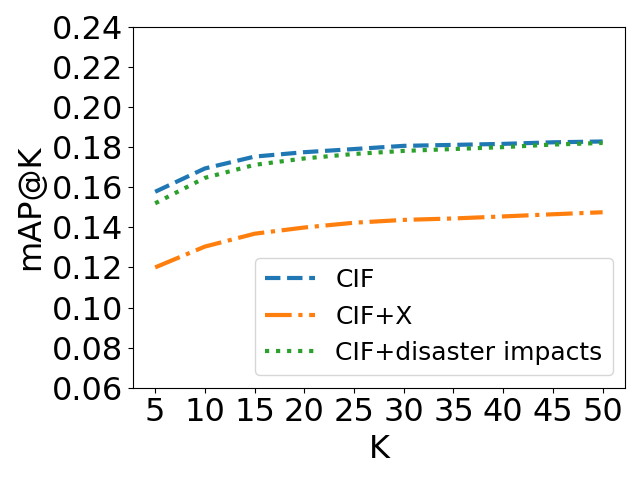}
    \caption{0h-24h}
    \end{subfigure}
    \caption{Performance comparison of retrieval queries for Christchurch}
    \label{fig:christchurch_retrieval_mAP}%
\end{figure}

We also experimented with $K$ to assess its effect on the retrieval performance. That is, we ran queries for top-$K$ where we changed $K$ from 5 to 50 with a step size of 5. We then computed the overall $\mathrm{mAP}@K$ in each time interval using Equation~(\ref{MAP_eq}) to evaluate the overall performance of each query type. Figures~\ref{fig:broward_retrieval_mAP} and \ref{fig:christchurch_retrieval_mAP} show the performance plots for all query types across different $K$ values and time intervals for Broward County and Christchurch, respectively. The plots initially show an increasing trend in performance with increasing $K$ but the performance improvement generally slows down after $K=30$. For Broward County, \textit{``CIF + disaster impacts''} queries outperform the other queries by a big margin whereas for Christchurch \textit{``CIF''} queries hold a slight edge over the other queries in the six-hour time intervals but perform only on par with \textit{``CIF + disaster impacts''} in the 0h-24h time interval.

Tables~\ref{tab:broward_retrival} and \ref{tab:cristhchurch_retrival} summarize the $\mathrm{mAP}@K$ scores for $K=50$ for Broward County and Christchurch, respectively. Again we observe that the scores for the \textit{``CIF + disaster impacts''} queries are highest in all time intervals for Broward County whereas \textit{``CIF''} and \textit{``CIF + disaster impacts''} queries perform on par for Christchurch. However, it is important to note that, for both AOIs, \textit{``CIF + disaster impacts''} queries retrieve the highest number of relevant tweets among all query types in 0h-24h time interval, i.e., 534 out of 1205 relevant tweets in Broward County and 402 out of 728 relevant tweets in Christchurch. Therefore, we conclude that \textit{``CIF + disaster impacts''} query type yields the best overall retrieval performance, and hence, we use it as the default query type in the proposed pipeline.


\begin{table}[t]
\centering
\begin{threeparttable}
\caption{\label{tab:broward_retrival}Retrieval performance ($\mathbf{\mathrm{\textbf{mAP}}@50}$) of the Broward County queries}
\begin{tabular}{lrrrrrr}
\toprule
 & \multicolumn{6}{c}{Time Intervals} \\
\cmidrule(lr){2-6}
Queries                       & \multicolumn{1}{c}{0h--6h} & \multicolumn{1}{c}{6h--12h} & \multicolumn{1}{c}{12h--18h} & \multicolumn{1}{c}{18h--24h} & \multicolumn{1}{c}{0h--24h} & \multicolumn{1}{c}{Y} \\ \midrule
CIF                                          & \underline{0.102}	& \underline{0.098} &	\underline{0.103} &	\underline{0.095} &	\underline{0.099} & 303 \\
CIF + X & 0.089	&0.088	&0.087	&0.090	&0.089    & \underline{484} \\
CIF + ``disaster impacts''                       & \textbf{0.139} & \textbf{0.133}	& \textbf{0.133}	& \textbf{0.111}	& \textbf{0.129} & \textbf{534} \\
\bottomrule
\end{tabular}
\begin{tablenotes}
\small
\item X = flooded, submerged, damaged, destroyed, weakened, cracked, blocked, torn, power outage, ruptured, collapsed, failed, uprooted, eroded, burnt, washed away, slippery, displaced, disrupted
\item Y = Number of relevant tweets retrieved out of 1205 in 0-24h time interval
\end{tablenotes}
\end{threeparttable}
\end{table}


\begin{table}[t]
\centering
\begin{threeparttable}
\caption{\label{tab:cristhchurch_retrival}Retrieval performance ($\mathbf{\mathrm{\textbf{mAP}}@50}$) of the Christchurch queries}
\begin{tabular}{lrrrrrr}
\toprule
 & \multicolumn{6}{c}{Time Intervals} \\
\cmidrule(lr){2-6}
Queries                       & \multicolumn{1}{c}{0h--6h} & \multicolumn{1}{c}{6h--12h} & \multicolumn{1}{c}{12h--18h} & \multicolumn{1}{c}{18h--24h} & \multicolumn{1}{c}{0h--24h} & \multicolumn{1}{c}{Y} \\ \midrule
CIF                     & \textbf{0.198}	& \textbf{0.229}	& \textbf{0.172}	& \textbf{0.148}	& \textbf{0.183} & 288 \\
CIF + X                  & 0.156	&0.178	&0.131	&0.134	&0.148  & \underline{326} \\
CIF + ``disaster impacts''  & \underline{0.187}	& \underline{0.212}	& \underline{0.167}	& \underline{0.169}	& \underline{0.182} & \textbf{402} \\
\bottomrule
\end{tabular}
\begin{tablenotes}
\small
\item X = flooded, destroyed, leak, blocked, cracked, ground liquefaction, power outage, ruptured, buried, collapsed, ground shake, unsafe, muddy
\item Y = Number of relevant tweets retrieved out of 728 in 0-24h time interval
\end{tablenotes}
\end{threeparttable}
\end{table}



\begin{figure}[t]%
    \centering
    \begin{subfigure}[b]{0.49\linewidth}
        \centering
        \includegraphics[width=\linewidth]{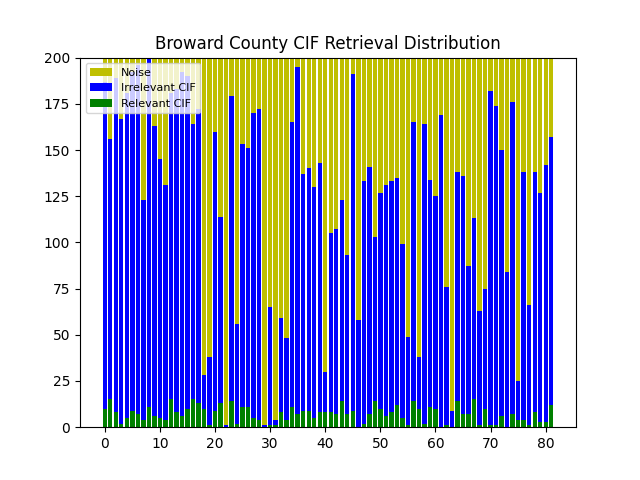}
        \caption{Broward County}
    \end{subfigure}
    \hfill
    \begin{subfigure}[b]{0.49\linewidth}
        \includegraphics[width=\linewidth]{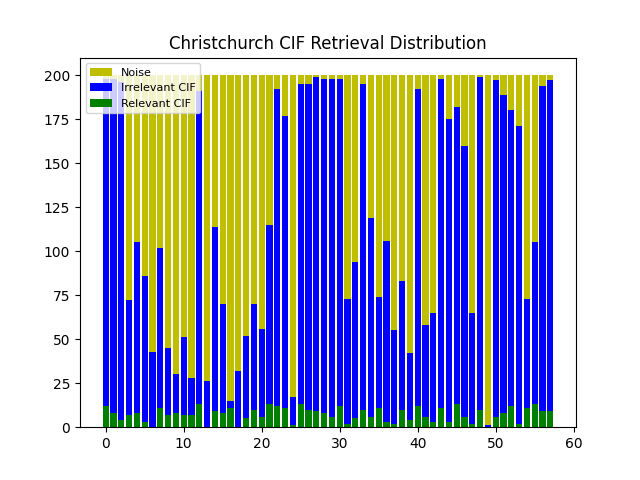}    
        \caption{Christchurch}
    \end{subfigure}
    \caption{Signal-to-noise distribution of retrieved tweets for each CIF in (a) Broward County and (b) Christchurch}%
    \label{fig:retrieval_distribution}%
\end{figure}

We note that the resulting mAP scores are generally low and the effective hit rate (i.e., ratio of relevant retrieved tweets to total number of retrieved tweets) is only slightly above 3\% for both AOIs. To investigate this further, we color coded and plotted the distribution of relevant and irrelevant tweets in Figure~\ref{fig:retrieval_distribution}. We additionally sub-divided irrelevant tweets belonging to either noise or other CIF that we generated. 
From Figure~\ref{fig:retrieval_distribution}, we observe that although a large portion of the distribution for each CIF contains irrelevant tweets, a significant portion of irrelevant tweets is occupied by tweets describing impacts to other CIFs and share some semantic similarity with the relevant tweet. For each query on average roughly 60\% of the retrieved results belong to a certain CIF (not necessarily the correct one) and the remaining contains pure noise. Although achieving a higher hit rate is always desirable, we defer further improvements for the retrieval process to our future work.

\subsection{Tweet Classification for Impact, Severity, and Operational Status}

Here we compare the ground-truth impact, severity, and operational status labels obtained from the Llama-2 13B model in the dataset generation stage with the labels inferred by the Mistral 7B v1.0 model in the data analysis and classification stage. Because we do not expect the classification models to depend on a specific CIF or time interval, we evaluate the classification performance over all tweets across all CIFs and time intervals using standard evaluation metrics such as Precision, Recall, and F1-score. Table~\ref{tab:classification_results_generated_data} summarizes the results of tweet-level impact, severity, and operational status classification for both AOIs. The results indicate that the proposed LLM-based classification models achieve reasonable performance for the task at hand. In general, the models achieve better precision than recall in all three tasks. However, the differences between precision and recall are bigger in the operational status classification task, which leads to the lowest F1-scores among all the tasks.

\begin{table}[t]
\caption{\label{tab:classification_results_generated_data}Performance scores for impact, severity, and operational status classification on the synthetic data}
\setlength\tabcolsep{4pt} 
\centering
\begin{tabular}{lccccccccc}
\toprule
         & \multicolumn{3}{c}{Impact} & \multicolumn{3}{c}{Severity} & \multicolumn{3}{c}{Operational Status}\\
         \cmidrule(lr){2-4}\cmidrule(lr){5-7}\cmidrule(lr){8-10}
AOI & Precision & Recall & F1-score & Precision & Recall & F1-score & Precision & Recall & F1-score \\
\midrule
Broward County &0.660 &0.563 &0.565             &0.668 &0.512 &0.566               &0.805& 0.349 &0.445                         \\
Christchurch   &0.603 &0.570 &0.569             &0.550 &0.458 &0.490               &0.655 &0.280 &0.342                         \\
\bottomrule
\end{tabular}
\end{table}

\begin{figure}[t]
    \centering

        \includegraphics[width=\linewidth]{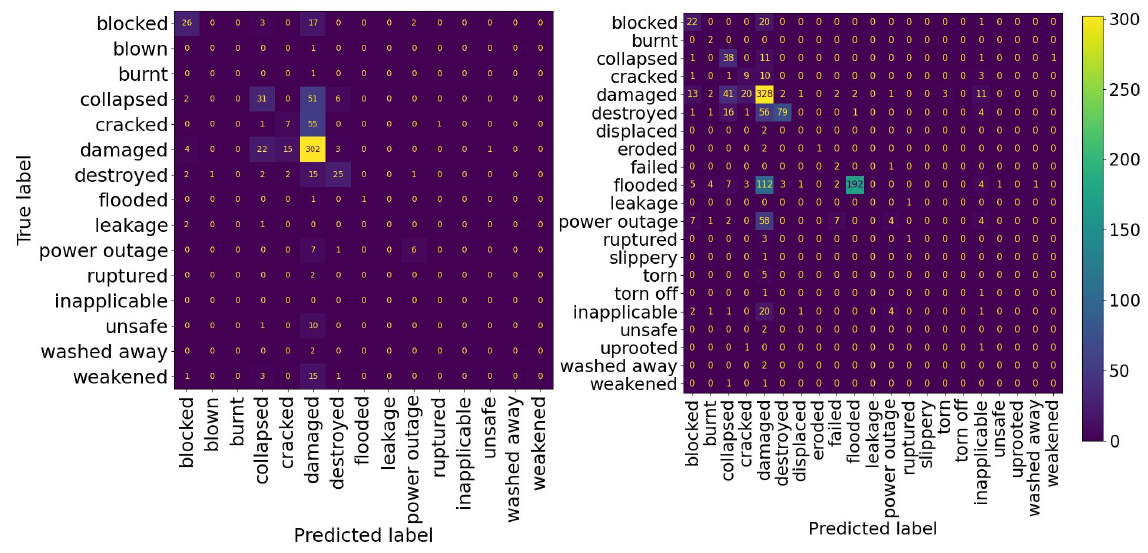}

    \caption{Confusion matrices for all signal impact classification (left) Broward County and (right) Christchurch}%
    \label{fig:impacts_confusion_matrix_generated_data}%
\end{figure}

To take a closer look at the impact classification performance on the synthetic data, we provided confusion matrices in Figure~\ref{fig:impacts_confusion_matrix_generated_data}. For Broward County, the most prominent impact classes such as \textit{damaged} and \textit{flooded} are correctly classified most of the time, but it is notable that around one third of the \textit{flooded} tweets are classified as \textit{damaged} (N=112). This is understandable because \textit{damaged} class has a broader meaning and coverage that overlaps with many other impact types including \textit{flooded}. Similarly, \textit{blocked} (N=20), \textit{destroyed} (N=56), and \textit{power outage} (N=58) tweets are usually classified as \textit{damaged}. However, the opposite is not correct, and in line with this expectation, we see much fewer \textit{damaged} tweets getting labeled as \textit{flooded} (N=2), \textit{blocked} (N=13), \textit{destroyed} (N=2), and \textit{power outage} (N=1). We also observe that two thirds of the \textit{unknown/inapplicable} tweets (N=20) are confused as \textit{damaged}, which is not very desirable and needs closer investigation. For Christchurch, we observe similar trends where, as the most prominent impact class, \textit{damaged} tweets are correctly classified most of the time (N=302), but many other impact tweets such as \textit{blocked} (N=17), \textit{collapsed} (N=51), \textit{cracked} (N=55), \textit{destroyed} (N=15), and \textit{weakened} (N=15) are also tagged as \textit{damaged}. Moreover, majority of the \textit{unknown/inapplicable} tweets (N=23) are confused as \textit{damaged} tweets. Altogether these observations suggest that it is challenging to get the Mistral 7B v1.0 model (or any other LLM for that matter) distinguish between fine-grained impact types.

We then also analyze the impact classification performance on the output of the retrieval system which contains noisy data due to imperfect nature of the retrieval process. Here we assume that ground-truth labels for the noise tweets belong to \textit{unknown/inapplicable} class for all classification tasks. Table~\ref{tab:classification_results_retrieved_data} summarizes the scores for impact, severity, and operational status classification on the retrieved tweets. These scores are higher thanks to the model's ability to correctly identify the (incorrectly retrieved) noise tweets as ``unknown/inapplicable.''

\begin{table}[t]
\caption{\label{tab:classification_results_retrieved_data}Performance scores for impact, severity, and operational status classification on the retrieved tweets}
\setlength\tabcolsep{4pt} 
\centering
\begin{tabular}{lccccccccc}
\toprule
         & \multicolumn{3}{c}{Impact} & \multicolumn{3}{c}{Severity} & \multicolumn{3}{c}{Operational Status}\\
         \cmidrule(lr){2-4}\cmidrule(lr){5-7}\cmidrule(lr){8-10}
AOI & Precision & Recall & F1-score & Precision & Recall & F1-score & Precision & Recall & F1-score \\
\midrule
Broward County &0.821 &0.747 & 0.769            &0.687 &0.6445 & 0.6442              &0.517 &0.352 & 0.399                        \\
Christchurch   &0.841 &0.805 & 0.777            &0.817 &0.794 & 0.797              &0.919 &0.779 & 0.809                        \\
\bottomrule
\end{tabular}
\end{table}

\begin{figure}[t]
    \centering
    \includegraphics[width=\linewidth]{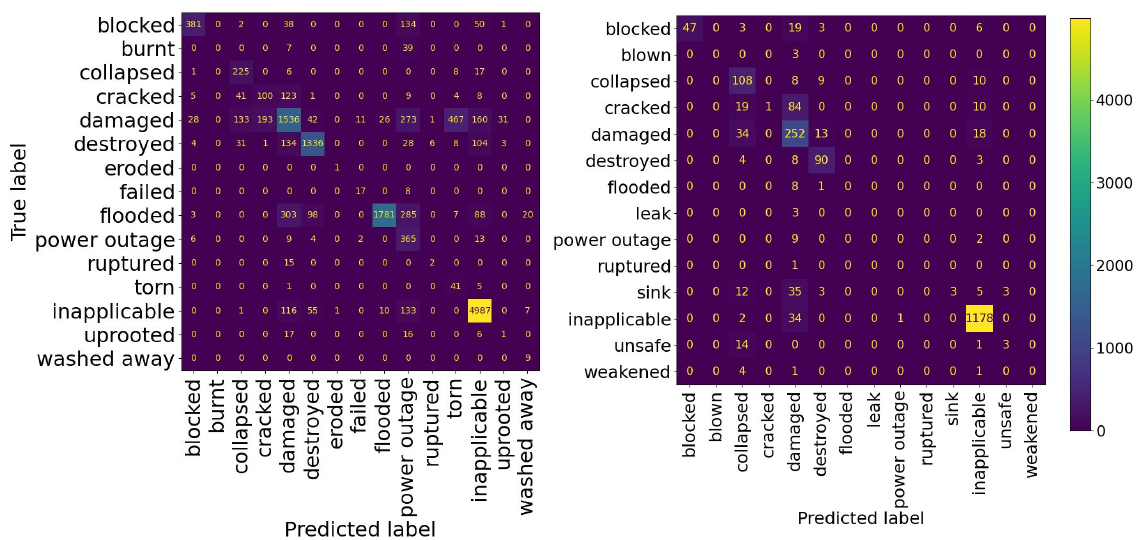}
    \caption{Confusion matrices for impact classification of retrieved tweets (left) Broward County and (right) Christchurch}%
    \label{fig:impacts_confusion_matrix_retrieved_data}%
\end{figure}

Figure~\ref{fig:impacts_confusion_matrix_retrieved_data} visualizes confusion matrices for further analysis of the classification results of the retrieved tweets. For both AOIs, the confusion matrices are dominated by the correctly identified noise tweets as \textit{unknown/inapplicable} (i.e., N=4987 for Broward County and N=1178 for Christchurch). Besides, for Broward County, majority of the tweets are correctly classified as \textit{damaged} (N=1536), \textit{destroyed} (N=1336), \textit{flooded} (N=1781), \textit{blocked} (N=381), and \textit{power outage} (N=365). However, many ground-truth \textit{damaged} tweets were also labeled incorrectly as \textit{torn} (N=467) or \textit{power outage} (N=273) and many \textit{flooded} tweets as \textit{damaged} (N=308) or \textit{power outage} (N=285). Whereas for Christchurch, while \textit{damaged} (N=252), \textit{destroyed} (N=90), \textit{collapsed} (N=108), and \textit{blocked} (N=47) tweets are classified correctly in general, we see that most \textit{cracked} tweets were confused as \textit{damaged} (N=84) and \textit{damaged} tweets as \textit{collapsed} (N=34). We also see that impacts such as \textit{burnt, ruptured, uprooted, sink,} and \textit{weakened} are oftentimes misclassified. Another interesting observation is that so many tweets are classified (incorrectly more than half of the time) as \textit{power outage} in Broward County whereas almost no tweet gets classified as \textit{power outage} in Christchurch.

\subsection{Overall Operational Status Classification of CIFs}

Tweets retrieved with the impact labels were processed using Mistral 7B v1.0 to infer the overall operational status of a target CIF. 
The results are provided in Table \ref{tab:classification_results_overall_status}. While exhibiting a reasonable precision, the model struggled with the recall. This could be due to the retrieval process, which incorporates messages from other CIFs and noise. Even though we mitigated the noise by focusing solely on impact-related messages, messages about other pertinent CIFs still pose a challenge for the model.

\begin{table}[t]
\caption{\label{tab:classification_results_overall_status}Performance scores for overall operational status classification on the retrieved results}
\setlength\tabcolsep{4pt} 
\centering
\begin{tabular}{lccc}
\toprule
         & \multicolumn{3}{c}{Overall Operational Status} \\
         \cmidrule(lr){2-4}
AOI & Precision & Recall & F1-score\\
\midrule
Broward County &0.522 &0.305 &0.216                                  \\
Christchurch   &0.470 &0.248 &0.197                                   \\
\bottomrule
\end{tabular}
\end{table}

\section{Discussion}
\label{sec:discussion}


Besides illustrating the opportunities LLMs have to offer for real-time monitoring of CIFs by analyzing social media data during disasters, this study also reveals various challenges to be tackled along the way. For instance, an important challenge is how to instruct LLMs to generate a realistic timeline of short messages (i.e., tweets) about CIFs during disasters that are coherent and consistent both spatially and temporally. Another important challenge is how to guide LLMs to eliminate irrelevant messages successfully in the retrieval process. Yet another important challenge is how to prompt LLMs in a zero-shot setting to discern fine-grained details about disaster impact types and severity affecting CIFs as well as to infer the operational status of CIFs from a set of impact and severity reports.

On the data side, we opted for open-ended generation of impact types. This allowed us to generate a quite diverse set of fine-grained impact types (N=83), a property sought by stakeholders to make key decisions. However, this strategy also resulted in a long-tail distribution of impact types where the generic \textit{damaged} became the most prominent impact type by a large margin while more than two thirds of the impact types appeared only once or twice in the entire dataset (see Figure~\ref{fig:impact_frequency_distribution}). An alternative strategy could be to impose more control over the distribution and ask LLM to generate tweets for a specific impact type at a time (potentially at the expense of reduced diversity).

On the retrieval side, our current approach experienced difficulty in eliminating the irrelevant content for some CIFs more than others (see Figure~\ref{fig:retrieval_distribution}). This could be attributed partly confusing nature of similar CIF or partial/shortened CIF names and partly to weaknesses of the retrieval method employed. In future studies, we plan to explore more sophisticated retrieval techniques.

On the classification side, we observed weaknesses of LLMs in distinguishing granular details about disaster impact types as analyzed through confusion matrices in Figures~\ref{fig:impacts_confusion_matrix_generated_data} and \ref{fig:impacts_confusion_matrix_retrieved_data}. There is also the fact that there are multiple correct answers many times and formulating the problem as a single-label classification becomes problematic as illustrated by the examples below:


\begin{quoting}
{\small{\it{
Case 1: The Category 5 hurricane has caused significant damage to the South Area Alternative School's electrical system, leaving the building without power. The school's future is uncertain.

Llama-2 13B (ground truth): Power outage\\
Mistral 7B v1.0  (classified): Damaged

Case 2: The hurricane has caused a tree to fall on top of a building near Broward Health Imperial Point, causing significant damage and blocking access to the hospital. 

Llama-2 13B (ground truth): Damaged\\
Mistral 7B v1.0  (classified): Blocked

Case 3: The University of Florida Field Laboratory's interior is flooded, causing significant damage to research equipment and infrastructure. The facility is closed indefinitely. 

Llama-2 13B (ground truth): Flooded\\
Mistral 7B v1.0  (classified): Damaged

Case 4: Honey Hill Fire Station's front entrance crushed by fallen tree. Emergency responders working to clear debris and reach those trapped. 

Llama-2 13B (ground truth): Collapsed\\
Mistral 7B v1.0  (classified): Blocked
}}}
\end{quoting}

In all cases, both ground-truth and classified labels seem reasonable with subtle differences stemming from how each LLM interprets the overall situation. 


Among all the tasks performed by LLMs, the most challenging proved to be inferring the overall operational status of a CIF. Given that the retrieval process may include messages that are semantically akin to a CIF query, potentially encompassing both, messages about other CIFs and noise, the model tends to face confusion when deducing the overall operational status. We remark that enhancing the retrieval process will significantly improve the performance of the inference task.

\section{Limitations and Future Work}
\label{sec:limitations}

In this section, we describe the biases and limitations of this study. Our main source of bias comes from the dataset utilized in this study. Due to the unavailability of real-world curated datasets describing impacts to CIFs, we utilized Llama-2 13B to generate the relevant tweets as well as the matching labels. In Figure \ref{fig:impact_frequency_distribution}, we note that the distributions of impacts generated by Llama-2 13B mainly cover broader impacts such as \textit{damaged, destroyed, flooded and collapsed}. Although Llama-2 13B also generated fine-grained impacts such as  \textit{shattered windows, ruptured pipe, HVAC damage and muddy road etc}, such distribution might not be representative of the distribution of impacts in real-world disaster. Likewise, the class/label imbalance in the dataset also limits us from fully accessing the capabilities of Mistral 7B v1.0 for the classification of various low-frequency classes. Moreover, the progression of impacts to CIFs in 24-hour time window was done by manually assigning the precedence of impacts to a CIF by the authors. This in part is also a bias that could have affected the prediction of the overall operational status performed in this study. 

Among the limitations, the most predominant ones are the language usage, the choice of LLM models and the nature of disasters selected for this study. Unlike real-world tweets, the synthetic tweets generated by Llama-2 13B were free from any typos, slang or grammatical errors. This study, therefore, does not fully assess the sensitivity of LLM-based classification to such inconsistencies. Moreover, we limited the type of disasters as well the choice of AOI used in this study to only two. This significantly limits the vocabulary of impacts as well as the nature of the content of the tweets in the generated dataset. The choice of AOI used, i.e., Christchurch and Broward County, also resulted in the generation and classification of only English medium tweets. 

Likewise, this study is also limited to one particular LLM for the data generation i.e., Llama-2 13B and another LLM for the classification task i.e., Mistral 7B v1.0. Therefore, further analysis of the quality of data generation and classification of crisis-relevant tweets by other LLMs is deferred to our future work.

\section{Conclusion}
\label{sec:conclusion}

In this study, we investigated the application of Large Language Models to monitor the operational status of critical facilities in an area during large-scale disasters. We utilized social sensing data (both synthetic and real-world) from X (formerly Twitter) for two distinct locations and performed their analysis using LLMs for various computational tasks, including retrieval, classification, and inference. The outcomes are presented using standard evaluation metrics and further discussed to highlight the strengths and weaknesses of LLMs. In summary, we emphasize that the use of LLMs in the zero-shot setting, as compared to supervised models requiring training data, demonstrates reasonable performance. However, challenges arise for LLMs when dealing with complex contextual information provided through lengthy prompts. Additionally, we outline various future directions that we anticipate to be beneficial for the community to build upon.




\printbibliography[heading=bibliography]









  









\end{document}